\documentclass{jetpl}
\twocolumn

\usepackage{hyperref}
\usepackage[numbers,sort&compress]{natbib}

\lat

\def \be {\begin{equation}}
\def \ee {\end{equation}}

\newcommand{\aleq}[1]{
\begin{equation}
    \begin{aligned}
    #1
    \end{aligned}
\end{equation}
}

\title{Effective Hamiltonian of topologically protected qubit in a helical crystal}

\rtitle{Effective Hamiltonian of topologically \ldots}

\sodtitle{Effective Hamiltonian of topologically protected qubit in a helical crystal}

\author{R.\,A.\,Niyazov$^{+*}$\/\thanks{e-mail: r.niyazov@mail.ioffe.ru}, 
D.\,N.\,Aristov$^{+*\times}$, 
V.\,Yu.\,Kachorovskii$^{*}$}

\rauthor{R.\,A.\,Niyazov, D.\,N.\,Aristov, V.\,Yu.\,Kachorovskii}

\sodauthor{Niyazov, Aristov, Kachorovskii}

\address{$^+$NRC ``Kurchatov Institute'', Petersburg Nuclear Physics Institute, Gatchina 188300, Russia\\~\\
$^*$Ioffe Institute, 194021 St.~Petersburg, Russia\\~\\
$^\times$Department of Physics, St. Petersburg State University, St. Petersburg 199034, Russia}

\dates{18 July 2023}{*}

\abstract{
We study a superlattice formed by tunnel-coupled identical antidots periodically situated in a two-dimensional topological insulator placed in a magnetic field. The superlattice spectrum can be controlled by gate electrodes or by changing the magnetic flux through the antidots.  We demonstrate that  a topologically protected qubit appears at the boundary between two regions with different fluxes. The qubit  properties depend on the  value of the flux jump on the boundary  and can be controlled by the gate voltage.  We derive the effective Hamiltonian of such a qubit  and analyze  the dependence of its properties on the main parameters of the superlattice: the tunnel coupling between antidots, and the probability of jumps with the spin flip.
 }

\begin{document}

\maketitle

{\bf Introduction.}
In the last decade, the electrical and optical properties of topological insulators, which are materials    insulating  in the bulk and possessing conducting states at the boundary, have been actively discussed~\cite{Bernevig2013,Hasan2010,Qi2011}. The properties of these states depend on the dimensionality of the material. In two-dimensional topological insulators, there are helical edge states (HES) at the edge that carry a non-dissipative current. Such one-dimensional states are characterized by a certain chirality, i.e. electrons with opposite spin propagate in opposite directions. Conventional (non-magnetic) disorder cannot reverse the spin of an electron, and precisely because of this property, there is no backscattering and dissipation in such channels.

The best-known implementation of 2D topological insulators is  HgTe-based quantum wells, whose topological properties have been theoretically predicted \cite{Kane2005,Bernevig2006} and confirmed by a series of experiments, including edge state conductivity measurements \cite{Konig2007} and experimental proof of non-local transport \cite{Roth2009,Gusev2011,Bruene2012,Kononov2015}.

One of the new and promising areas of research is interferometry based on HES. The most interesting effects  arise in the presence of a magnetic field, which adds additional physics to the problem by controlling interference through the Aharonov-Bohm (AB) effect. This effect manifests itself in the bulk properties of two-dimensional topological insulators \cite{Peng2010,Lin2017,Bardarson2010,Bardarson2013, Gusev2015}, as well as in systems with HES and interferometers based on them \cite{Delplace2012,Dolcini2011,Gusev2015,Niyazov2018,Niyazov2020,Niyazov2021,Niyazov2021a}.

Usually, the interference is suppressed when the temperature, $T,$ becomes larger than the distance between quantization levels in the system. As was  theoretically demonstrated recently \cite{Niyazov2018, Niyazov2020,Niyazov2021,Niyazov2021a}, this is not the case for AB interferometers based on HES, where the interference is preserved even for the case $T\gg 2\pi v_{\rm F}/ (L_1+ L_2)$ (here $L_{1,2}$ are the lengths of the interferometer arms, and $v_{\rm F}$ is the Fermi velocity). This means that interference effects in HES-based systems can be studied at relatively high temperatures, which are practically important for various applications.

One of the promising directions for further research is the study of HES-based metamaterials, such as helical crystals.  A helical crystal can be created by chemically etching holes through a layer of two-dimensional topological material. At the edges of such holes, the HES appear, which are  connected by tunneling or electrically.
The first experimental realizations of systems of periodically arranged holes in HgTe/CdTe based structures   \cite{Maier2017,Ziegler2019} have already been reported. The magnetotransport properties of periodic arrays of such holes were studied.  However, the distance between holes was sufficiently large so that tunneling between HES was impossible even in the topological phase.
Similar structures with shorter spacing between etched holes, and therefore stronger tunneling  coupling between HES corresponding to adjacent holes, could exhibit tunneling. Moreover, tunnel barriers between holes can be controlled using gate electrodes. Thus, it is possible to produce a one- or two-dimensional array of tunnel-coupled  HES.  Such a system
is a   realization of the helical crystal.

The simplest implementation of such a crystal, namely a one-dimensional periodic array of closely spaced identical etched holes  was recently considered theoretically in our work \cite{Niyazov2023}. We have shown that the band structure of a crystal can be controlled both by gate electrodes and by an external magnetic field by changing the magnetic flux through the antidot.

\begin{figure*}
\includegraphics[width=0.97\linewidth]{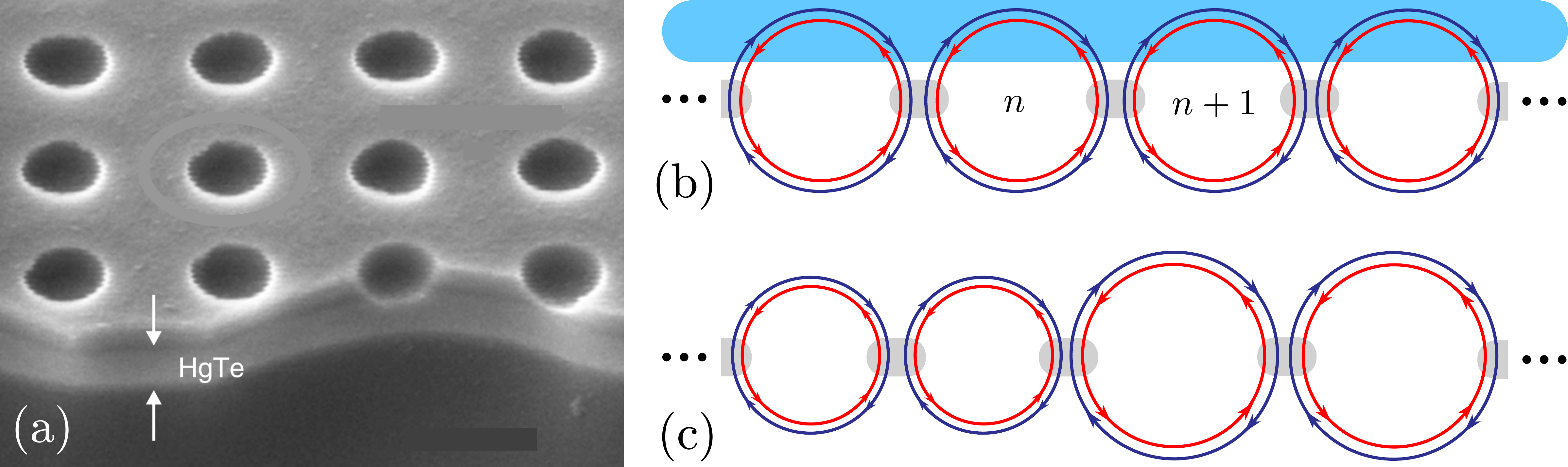} 
\caption{\label{fig:1Crystal}
(a) An example of an experimental implementation of a helical crystal, taken from~\cite{Maier2017};
(b) One-dimensional chain of helical rings formed by HES (marked in red and blue) connected by tunnel or ballistic contacts (marked in grey). The control gate is attached to the top of the crystal (indicated in light blue);
(c) The boundary between two semi-infinite  arrays of rings with different radii, on which a flux jump occurs, and, as a consequence, a localized qubit appears.
}
\end{figure*}

It has been demonstrated that changing the magnetic field makes it possible to create Dirac points in the spectrum. Such points appear at special field values corresponding to an integer or half-integer quantum of the magnetic flux through the hole.
The deviation of the flow from special values leads to the appearance of gaps, i.e. generation of a finite mass of Dirac fermions.
  In such a crystal, various kinds of defects can be created, for example, by inserting one antidot of a different size into the crystal, through which a magnetic flux passes, different from the flow through other points. It is also possible to connect two regions with different sizes of antidots to create a flux jump at the boundary (see Fig.~\ref{fig:1Crystal}). Also, a flux jump (FJ) can be created by applying an inhomogeneous magnetic field to an array of identical antidots. As we will show below, the FJ leads to the appearance of doubly degenerate topologically protected states of the Volkov-Pankratov type~\cite{Volkov1985}.
The degeneracy can be removed by changing the magnetic field or, more importantly, by purely electrical means by changing the gate potentials. This means the formation of a topologically protected qubit that can be manipulated using gate electrodes.

The goal of this work is the derivation of the effective Hamiltonian describing such a qubit and to analyze the dependence of the properties of the qubit on  key parameters of the superlattice: the tunnel coupling between antidots and the probability of spin-flip tunneling.

{\bf Model.}
We study a one-dimensional array of antidots (etched holes) on a surface of the two-dimensional spin-Hall insulator [see fig.~\ref{fig:1Crystal}~(a)].
There are HES at the boundaries of the holes, generally forming a chain of identical helical rings with circumference $L=L_1+L_2$, connected in series [see fig.~\ref{fig:1Crystal}~(b)]. For simplicity, we  assume below $L_1=L_2$. Each ring contains electrons moving clockwise and counterclockwise with energy $E,$ and momentum $k= E/v_{\rm F},$ with opposite spins at each point. A uniform magnetic field is applied to the array, which creates a $\Phi$ flux in each ring. In this paper, we restrict ourselves to the consideration of a non-interacting system. Interesting problems that arise due to the electron-electron interaction will  briefly be discussed in the conclusions part of the paper.

Let us assume that the $X$-type junction between two adjacent rings with four scattering channels
is described by a general scattering matrix (up to insignificant phases) $\hat S$, which preserves symmetry under time reversal~\cite{Teo2009,Aristov2016}:
\be
C_\alpha^\prime= S_{\alpha \beta} C_\beta, \quad \alpha,\beta=1,2,3,4;
\ee 
\aleq{ \label{eq:smat}
\hat S&=\begin{pmatrix}
0 & t & f & i r ^* \\
t & 0 &  i r    & f \\
-f &  i r   & 0 &   t \\
 i r ^*& -f &   t & 0  \\
\end{pmatrix} \,. 
}
Here $C_\alpha^\prime$ and $C_\beta$ are the amplitudes of the outgoing and incoming waves, respectively, and the real-valued amplitudes
$t=\cos \alpha$, $f=\sin \alpha \sin \beta$, $r= \sin \alpha \cos \beta e^{i v/2}$ satisfy the condition
$t^2+|r|^2+f^2=1.$ Strictly speaking, the amplitudes $t$ and $f$ may be complex-valued, but the phases of these coefficients  do not enter the dispersion relations discussed below and can be omitted, see discussion in \cite{Niyazov2023}.

Physically, the amplitudes $r$ and $f$ describe jumps between rings  with the same and opposite spin projections, respectively. The parameter $\alpha$ is related to the total probability of such a jump: $|r|^2+f^2=\sin^2 \alpha.$ In other words, the angle $\alpha$ is responsible for the tunnel connection between the rings, and  the rings become independent at $\alpha=0$. For a tunnel contact ($\alpha \ll 1$), the value of $\alpha$ is mainly determined by the tunneling distance. The case of ballistic (or metallic) contact, when the rings are strongly coupled, corresponds to $\alpha \approx \pi/2.$
The angle $2\beta$ is the angle between the spin quantization axes in different helical states~\cite{Aristov2017}, and the elements $r$ and $f$ are proportional to the overlap of spinors in the HES in neighboring antidots. Accordingly, $\beta$ is responsible for the probability of spin-flip jumps between rings. For $\beta=0, $ the $f$ amplitude  vanishes and, as a consequence, there are no spin-flip processes.

The phase $v$ is a key  parameter which  determines the qualitative properties of the spectrum.  Remarkably,   this parameter can be controlled in a purely electrical way by changing the gate voltage \cite{Niyazov2023}.
 
Next, we introduce the transfer matrix, $\hat T$,   describing the transition from the $n-$th to the $(n+1)-$th cell of the helical crystal:
\be 
\hat T =  \hat P \hat T_0 .
\label{T}
\ee 
Here $T_0$ and $P$ describe transitions through the contact and the ring, respectively. The transfer matrix, $T_0$,
corresponds to the matrix $\hat S$ and expresses the amplitudes of the wave function in the $(n+1)-$th ring (immediately after the contact) in terms of the amplitudes in the $n-$th ring (immediately before the contact).
The matrix
\aleq{
\hat P=\text{diag} [e^{i \varphi_b},e^{-i \varphi_a},e^{i \varphi_a},e^{-i \varphi_b}]\, ,
\label{def:Pmat}
}
contains phases acquired inside the ring during the transition from the left contact to the right one:
\be
\varphi_a= \frac{\varepsilon}{2} - \pi \phi, \quad
\varphi_b= \frac{\varepsilon}{2} + \pi \phi.
\label{phases}
\ee
Here we have introduced dimensionless energy and dimensionless flow:
\be
\varepsilon= \frac{E L}{v_{\rm F} }= k L, \qquad   \phi=\frac{\Phi}{\Phi_0},
\label{dimen-E-Phi}
\ee
with  $\Phi_0$  the flux quantum.

{\bf General dispersion relation.}
The band structure of the crystal is determined by the following relation connecting the amplitudes on the $n$th ring to the amplitudes on the $(n+1)$th ring
$ \psi_{n+1} = \hat T \psi_{n} = e^{iq} \psi_n$. It leads to the equation:
\aleq{\label{eq:dispeq}
\det [\hat 1 e^{iq} -\hat T]=0 \,,
}
where $\hat 1$ is the $4\times 4$ identity matrix. The resulting dispersion relation $\varepsilon(q)$ can be found in an implicit form:
\aleq{
&\left(\cos \beta  \sin \left(  {\varepsilon }/2+\pi  \phi \right)+\sin \alpha  \cos (q- v/2)\right) \\
&\times \left(\cos \beta  \sin \left( {\varepsilon }/2-\pi  \phi \right) +\sin \alpha  \cos (q+ v/2 )\right)\\
&+\tfrac{1}{2} \sin ^2\beta  \left(\cos ^2\alpha  \cos 2 \pi  \phi -\sin ^2\alpha  \cos v -\cos \varepsilon \right)=0.    
\label{disp-general}
} 

{\bf Dirac points.}
\label{sec:Dirac-points}
The most interesting property of Eq.\ \eqref{disp-general} is the presence of Dirac points in the dependence $\varepsilon(q).$ Such points appear at integer and half-integer values of the dimensionless flux.
The Dirac points
appear in the vicinity of integer values $q/\pi$ for integer $\phi$ and near half-integer values $q/\pi$  for half-integer $\phi.$
Let us first discuss the spectrum in case of the flux close to integer values.
The position of the Dirac points can be represented as:
$\varepsilon_{\rm DP}=2 \pi n - 2(-1)^{n+m+p} \Theta,~ \phi_\text{DP} =p, ~
q_\text{DP} = \pi m .$
Here $n,$ $p$ are integers, $m = (-1,\,0) $, and $\Theta= \arcsin[\sin \alpha \cos \beta \cos(v/2) ].$ For the weak tunnelling $\Theta$ is small, $\Theta \propto \alpha \ll 1.$

In the vicinity of  Dirac points, the spectrum can be written as \cite{Niyazov2023}: 
  \be
  \delta \varepsilon =  \pm
  \sqrt{(A \delta q + B \delta    \phi)^2 + C^2  (\delta  \phi) ^2   },
  \label{spect}
  \ee
where $\delta \varepsilon,$ $\delta q,$ and $\delta\phi$ are measured from $\varepsilon_{\rm DP}, ~q_{\rm DP}$ and $ \phi_{\rm DP} , $ respectively.
  As follows from this equation, the gap in the spectrum is determined by the expression
 \be
 \Delta_{\rm DP}= 2 C |\delta \phi|.
 \ee
The coefficients $B$ and $C$ are the same for all Dirac points, while $A$ has a different sign but the same absolute values:
  $  A = 2~(-1)^{n+m+p}   \sin \alpha \sin
  \xi/\cos\Theta,$  $B= { 2\pi \cos\beta \sin(v/2) }/\sin \xi ,$  $ C= { 2 \pi \cos\alpha \sin \beta }/({\sin \xi \cos \Theta}),
 $ and $\xi= \arccos[\cos\beta \cos(v/2)]. $
  
Note that $A\to 0$ for $\alpha \to 0,$ thus reflecting the existence of flat bands in this limit. The notion of Dirac points is meaningful if the bandwidth is greater than the Dirac gap, $A \gg \Delta_{\rm DP}.$
  
For $\beta=0$ the helical channels are not connected. In this case $C=0$ and the spectrum becomes
  \be
  \delta \varepsilon = |A \delta q+ B \delta \phi|.
  \ee
Therefore, in this particular case, we have a gapless spectrum for any value of $\delta \phi.$

The spectrum for the case of almost half-integer values of the flux $\phi$ has
another set of Dirac points:
 $
\varepsilon_\text{DP}=2 \pi (n+\tfrac 12) - 2(-1)^{n+m+p} \Theta,~
\phi_\text{DP} =m+ 1/2,~ 
q_\text{DP}= \pi p +  \pi/2.$
Eq.\ ~\eqref{spect} remains true in this case with the replacement
  $v \to v+\pi$ in the coefficients $A,B,C$ and in the definitions of $\Theta$ and $\xi.$
  For small $v\ll 1$, the form of the coefficients in \eqref{spect} is simplified,
$  A \simeq 2~(-1)^{n+m+p}   \sin \alpha,$ 
$  B \simeq   2\pi \cos\beta ,$ $  C \simeq  { 2 \pi \cos\alpha \sin \beta } $, $2\Theta \simeq - v   \sin \alpha  \cos\beta.$
 
{\bf Effective Hamiltonian in the vicinity of  the Dirac point.}
One of  the easiest way to get the effective Hamiltonian near the Dirac points is to ``square''  the dispersion relation. Namely, instead of~\eqref{eq:dispeq} we consider the transition matrix to the second neighbor, $ \psi_{n+2} = \hat T^2 \psi_{n} = e^{2iq} \psi_n$, which leads to the equation
\aleq{
\det [\hat 1 e^{i2q} -\hat T^2(\varepsilon)]=0 \, . 
\label{TT}}
To be specific, consider $\phi\simeq 1/2,$ when the positions of the Dirac points are given by $q_{\rm DP} = \pm \pi/2 ,$ so that $ e^{2iq_ {\rm DP}} =- 1 .$ One can check that
$\hat T^2=-\hat 1$ also for $\varepsilon=\varepsilon_{\rm DP}$, although $\hat T$ is not diagonal. This property simplifies the expansion near the Dirac points: $q=\pm \pi /2+ \delta q$, $\varepsilon =\pi +\delta \varepsilon$, $\phi= 1/2 + \delta \phi$,
  $v=\delta v$ ($\delta \phi \ll 1, \delta q \ll 1, ~ \delta v \ll 1$).
The energy dependence in \eqref{TT} is encoded in the matrix $\hat P,$ which is included in the definition of $\hat T.$ Expanding this matrix  up to the second order of small $\delta \varepsilon$, we write
\aleq{
\hat P=\hat P_\pi (1-\hat I \delta \varepsilon/2-\delta \varepsilon^2/8) \,.
}
Here $\hat P_\pi=\hat P (\varepsilon=\pi)$ and $\hat I= \text{diag}[-i,i,-i,i]$.

In order to obtain the effective Hamiltonian describing a crystal in the vicinity of the Dirac point,  we first expand the expression in Eq.\ \eqref{TT}  up to the second order of $\delta \varepsilon$:
\aleq{
( \hat M_0 +\hat M_1 \delta \varepsilon +\hat M_2 \delta \varepsilon^2 ) \psi_n =0 \,,
\label{det-M}
}
where 
\aleq{
\hat M_0 &= (P_\pi T_0)^2-e^{2iq}\,, \\
\hat M_1 &= -\frac{1}{2} P_\pi( \hat T_0 \hat I + \hat I T_0 ) P_\pi T_0\,,\\
\hat M_2 &= \frac{1}{4} \hat P_\pi \hat I (T_0 \hat I + \hat I T_0) P_\pi T_0 \,.
}
In the derivation we used that $[\hat I, \hat P_{\pi}]=0.$
Eq. \eqref{det-M}  is rewritten as
\aleq{
 ( \hat 1 \delta \varepsilon  +\hat M_1^{-1}\hat M_0+ \hat M_1^{-1} \hat M_2 \delta \varepsilon^2 ) \psi_n =0 \,.
\label{de-M}
}
Multiplying from the left by $( \hat 1 - \hat M_1^{-1} \hat M_2 \delta \varepsilon ) $ and discarding the terms $\sim \delta \varepsilon^3$,
we get $( (  \hat 1 - \hat M_1^{-1} \hat M_2 \hat M_1^{-1}\hat M_0 )\delta \varepsilon + \hat M_1^{-1}\hat M_0 )  \psi_n $. 
This can be represented as the ``Schr\"odinger equation''
\aleq{
\delta \varepsilon \, \psi_n  & =  -  (  \hat 1 - \hat M_1^{-1} \hat M_2 \hat M_1^{-1}\hat M_0 )^{-1}   \hat M_1^{-1}\hat M_0 \psi_n 
\\ & \simeq  - (  \hat M_1^{-1}\hat M_0  + \hat M_1^{-1} \hat M_2 ( \hat M_1^{-1}\hat M_0 )^{2}  ) \psi_n 
\label{de1-M}
} 

The next step is to make an expansion with respect to  $\delta \phi,\delta q$ and $\delta v.$ As one can see, the matrix $\hat M_0 $ vanishes identically when $\delta \phi=\delta q= \delta v=0.$ Therefore, $\delta \varepsilon$ is small. In what follows, we will assume that $\delta \varepsilon \lesssim \delta \phi \sim \delta q \sim \delta v.$ Then the last term in \eqref{de1-M} is small and can be discarded within the lowest order. 
  
 \underline{ \emph{First order with respect to   $\delta \phi, \delta v, \delta q$ and $\delta \varepsilon$.}}
Let us first make the calculation keeping only the terms linear in $\delta \phi, \delta v, \delta q$, and $\delta \varepsilon$.
In this order, we have
\aleq{
 ( \hat 1 \delta \varepsilon  +\hat M_1^{-1}\hat M_0) \psi_n =0 \,.
\label{de-M1}
}
Note that the matrix $\hat M_1^{-1}\hat M_0$ is not Hermitian, but can be reduced to the Hermitian form by a certain  transformation $\hat R.$  The Hermitian operator arising after such a transformation     
\be
\mathcal{H}^{(1)}_\text{eff}=- \hat R^{-1}  \hat M_1^{-1}\hat M_0 \hat R,
\ee
has the same dispersion law (near the Dirac point) and has the meaning of the effective Hamiltonian.

The choice of the  matrix $\hat R$ is not unique is worth an additional comment. Omitting simple technical details, we will outline the general idea. First, we can find a matrix that diagonalizes $\hat M_1^{-1}\hat M_0$ in the simplest case $\delta \phi=0.$ In this case, the diagonalizing matrix $R_*$ can be multiplied from the right by the diagonal matrix  $ \hat C=\text{diag}[c_1,c_2,c_3,c_4],$ with arbitrary coefficients $c_i.$ Without loss of generality, we can demand $c_1 c_2 c_3 c_4 =1$ and choose $ \hat C$  in the  form $ \hat C = \text{diag}[a c, a c^{-1},a^{-1} c',a^{-1} c'^{-1}]$.
Now acting by the matrix $R_*$ on $\hat M_1^{-1}\hat M_0$ with $\delta \phi \neq 0$ we obtain a block matrix,
\aleq{\mathcal{H}^{(1)}_\text{eff} &=
\begin{pmatrix}
\mathcal{H}_1& 0\\
0& \mathcal{H}_2\\
\end{pmatrix}. }
The Hamiltonians $\mathcal{H}_{1,2}$ are off-diagonal (to the extent of $\delta \phi$) and, generally speaking, are not Hermitian, but can be reduced to Hermitian form for certain $c, c'$. Thus, the analysis within the framework of the lowest order leaves the coefficient $a$ arbitrary. As will be seen from the analysis of higher orders, it is natural to put $|a|=1$ and to choose the phase of this coefficient at our convenience.

The corresponding similarity matrix takes the form:      
\aleq{
R & =
\begin{pmatrix} 
 -i \tau ^{\alpha}\tau^\beta & i & - \tau ^{\beta } &   \tau ^{\alpha } \\
  1 & \tau ^{\alpha}\tau^\beta & - i \tau ^{\alpha } & -i\tau ^{\beta } \\
 \tau ^{\alpha } & \tau ^{\beta } & -i &  -i \tau ^{\alpha}\tau^\beta \\
 -i \tau ^{\beta } & i \tau ^{\alpha } & - \tau ^{\alpha }\tau^\beta & 1 
\end{pmatrix} \,,\\ 
\tau^\alpha & = \tan(\tfrac{\pi}{4}+\tfrac{\alpha}{2}) \,, \quad 
\tau^\beta   = \tan{\tfrac{\beta}{2}} \,,\\
}
and the blocks of the effective Hamiltonian become
\aleq{
\mathcal{H}_1 & =
\delta v \sin\alpha \cos \beta  \sigma_0 +2 \pi \delta \phi \cos \alpha \sin \beta \sigma_1
\\   & 
+ (2 \pi \delta \phi \cos \beta + 2 \delta q \sin \alpha) \sigma_3  \, , \\ 
\mathcal{H}_2 & =
-\delta v \sin\alpha \cos \beta  \sigma_0 +2 \pi \delta \phi \cos \alpha \sin \beta \sigma_1
\\ & 
+ (2 \pi \delta \phi \cos \beta  - 2 \delta q \sin \alpha) \sigma_3  \, . 
}
Here $\sigma_{1,3}$ are Pauli matrices and $\sigma_0$ is the identity matrix.
These formulas are consistent with Eq.\ \eqref{spect} above.

The block $\mathcal{H}_2$ of the Hamiltonian can be obtained from $\mathcal{H}_1$ by replacing $\delta q \to -\delta q$ and $\delta v \to -\delta v$ (or $\alpha \to-\alpha$).
The $\delta v$ term can be absorbed by the energy shift in each of  two blocks:
\be
  E_1= - E_2= -W,
\ee
where
\be 
W= \delta v \cos \beta \sin \alpha
\label{W}
\ee
Therefore, we can only discuss the  block $\mathcal{H}_1 $.
The small shift of the wave vector, $\delta q \ll 1 $, can be represented as the motion of an electron along an array of rings, $\delta q =-i d/dx,$ where $x$ is the coordinate along this array in units of the ring diameter, $L /\pi.$
By rescaling this dimensionless coordinate: $ {2} \sin \alpha\, \bar x = x ,$ we find
    ${2} \sin \alpha \delta q =
    -i \tfrac{\partial}{\partial \bar x}$. Finally, $\bar{\mathcal{H}}_1 $ becomes the Hamiltonian of the massive Dirac particle:
\aleq{
\bar{\mathcal{H}}_1=
\begin{pmatrix}
\bar \phi - i \frac{\partial}{\partial {\bar x}} & h \\
h & -\bar \phi + i \frac{\partial}{\partial {\bar x}}\\ 
\end{pmatrix} \,,
}
where $\bar\phi = 2\pi\delta\phi\cos\beta$ and $h= 2\pi\delta\phi\sin\beta\cos\alpha$. The dispersion of such Hamiltonian for a plane wave $e^{i\bar q {\bar x}}$ is given by $E^2=(\bar q+\bar \phi)^2+ h^2,$ where
\be 
\bar q= 2 \sin \alpha ~\delta q
=-i\frac{d}{d \bar x} \,.
\label{barq}
\ee

\noindent
 \underline{ \emph{Second order in  $\delta \phi, \delta v, \delta q$ and $\delta \varepsilon$.}}
Let us now expand Eq.\ \eqref{TT} up to the second order in $\delta \phi,~\delta q $ and $\delta \varepsilon.$ We use the following expressions:
\aleq{
\det [\delta \varepsilon + \hat M_1^{-1} \hat M_0+\hat M_1^{-1} \hat M_2 ( \hat M_1^{-1} \hat M_0)^2]=0 \,,
}
\aleq{
\mathcal{H}_\text{eff}=-R^{-1}( \hat M_1^{-1} \hat M_0+ \hat M_1^{-1} \hat M_2 (\hat M_1^{-1} \hat M_0)^2)R \,.
}
Corrections to ${\mathcal{H}}_{\rm eff}$ are of two types: (i) corrections to $ \bar{ \mathcal{ H}}_1$ and $\bar{ \mathcal{ H}}_2,$ which do not lead to anticrossing and can be omitted here, and (ii) block-off-diagonal corrections, which are given by the formula
 \begin{equation}
    \mathcal{H}^{(2)}_\text{eff} =
\begin{pmatrix}
0& \hat V  \\
\hat V ^\dagger& 0\\
\end{pmatrix} \,.\\
\end{equation}
With the accuracy described above, we find
\aleq{
\hat V &= 
  (\delta q^2 -  \pi^2 \delta\phi^2 \sin^2 \beta) \sin 2 \alpha\,\hat \sigma_3 
 \\
 &+   \pi^2 \delta \phi^2 \sin \alpha \sin 2 \beta \,\hat \sigma_1,
 \label{2nd-order}
}
where $\hat\sigma_i$ ($i=0,1,2,3$) are Pauli matrices.

{\bf Flux jumps in a crystal:
topologically protected localized states.}
Let us now discuss localized states near the FJ. We assume that the rings' array consists of two semi-infinite areas with different flux values [see fig.~\ref{fig:1Crystal}~(c)]. In the first region, at $x<0$, the flow is less than $1/2$, $\phi=1/2-\delta\phi$, and in the second region, at $x>0$, the flux exceeds 1/2, $\phi=1/2+ \delta\phi$.
The problem is similar to  one considered in the paper~\cite{Volkov1985}. Then, at the interface between the two regions, there appear two (since the Hamiltonian has two blocks) degenerate localized states of the Volkov-Pankratov type :
\aleq{
\psi^{\text{loc}}_1&= 
\begin{pmatrix}
i \\
1 \\
0\\
0\\
\end{pmatrix}
e^{-(|h| + i|\bar \phi |)|{\bar x}|}, \,  \\
\psi^{\text{loc}}_2&= 
\begin{pmatrix}
0\\
0 \\
1\\
i\\
\end{pmatrix}
e^{-(|h| -i|\bar \phi |)|{\bar x}|} \, 
\label{localized}
}
Notice that we have omitted the prefactor $(-1)^{n}=\exp( \pm i \pi x )$ in both functions, which corresponds to the shift $\delta q= q \pm \pi/2$ above.
We see that the localization lengths of these localized states are given by
\be L_{\text{loc}} = L ~\frac{ \tan \alpha }{\pi^2 \delta\phi \sin \beta }. 
\ee  In the derivation, we assumed that $L_{\text{loc}} \gg L$. For $\delta v=0,$ and $ \delta \phi \to 0 ,$ the energy corresponding to these degenerate states lies exactly in the middle of the Dirac gap, as in the problem discussed in~\cite{Volkov1985}.

The localized topologically protected states obtained above form a two-level system or qubit. The energy levels of this system
depend on the parameter $\delta v$, which controls the up-down asymmetry of the system.
By changing  $\delta v$ with the use of  gate electrodes, one can manipulate the energy levels $E_1=E_1(\delta v)$ and $E_2=E_2(\delta v)$, which intersect each other at $\delta v=0.$ At this point the levels are doubly degenerate in the lowest approximation in $\delta \phi$ and  $\delta q.$

In the next section we  show that, in fact,
there is anticrossing between these levels so there is ``level repulsion'' due to corrections to the effective Hamiltonian, see \eqref{2nd-order}.

\begin{figure}
\includegraphics[width=0.8\linewidth]{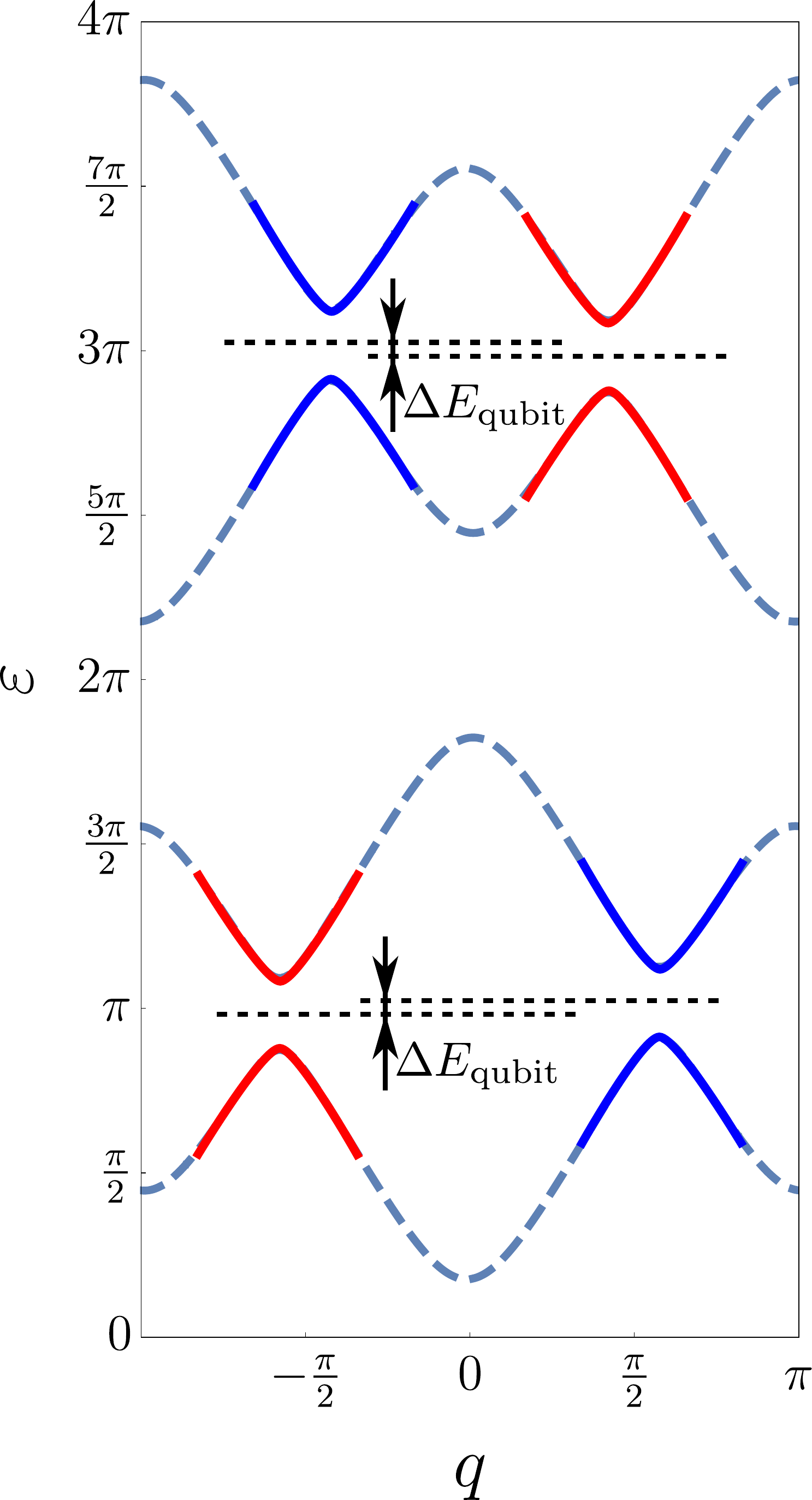} 
\caption{\label{fig:qubit}
Dispersion of low-lying bands of the helical crystal. The vertical distance between the Dirac points is controlled by  $v.$  The dotted lines show the position of the localized levels that arise in the presence of a flux jump in the crystal. $\Delta E_{\rm qubit}=E^+_{\rm qubit} -E^-_{\rm qubit}$, see Eq.\ ~\eqref{Epm-qubit}.
}
\end{figure}

{\bf Effective Hamiltonian of  qubit.}
Projecting $H_{\rm eff}^{(2)}$ onto two localized states, Eq.~\eqref{localized}, we get the Hamiltonian of the qubit formed by $\psi_1^{\rm loc}$ and$ \psi_2^{\rm loc}$:
\begin{equation}
 \mathcal{H}_{\rm qubit} =
\begin{pmatrix} W & \Delta  \\
\Delta^*  &  -W
\end{pmatrix} \,,
\label{Hq}
\end{equation}
Here $W$ depends linearly on $\delta v$ according to Eq.\ ~\eqref{W},
$\Delta = \Delta_x- i \Delta_y= 2\pi \delta \phi^2 \cot \alpha \sin\beta (i \cos \alpha \cos \beta -\sin\beta).$ 
The corresponding energy levels are given by
\be
 E^{\pm}_{\rm qubit}= \pm \sqrt{W^2 +  |\Delta|^2  },
\label{Epm-qubit}
\ee
where
\aleq{
|\Delta| & = 2\pi^2 \delta \phi^2 
\sin\beta \cos\alpha \sqrt{\cot^2\alpha + \sin^2 \beta} \,.
\label{D-qubit}
} 
The qubit levels are shown in Fig.~\ref{fig:qubit}. The distance between the levels depends on the tunnel coupling between the antidots, controlled by the parameter $\alpha,$ and on the probability of tunnel hopping accompanied by spin flip, which depends on $\beta.$  One  can  see that in the absence of spin-flip processes, i.e. for $\beta=0,$ we have $\Delta=0$. Notice also that in the limit of small $\alpha$ the expression \eqref{D-qubit} can be represented as
\aleq{
|\Delta| &  
\propto |\delta \phi| \, \Delta_{\rm DP} / A,
\label{D-qubit2}
} 
which shows that the stationary qubit exists at $|\delta \phi| <A \propto \alpha .$ In the opposite case,  $|\delta \phi| \geq A ,$ the qubit levels move into the continuous spectrum, $|\Delta| \geq \Delta_{\rm DP},$ i.e. they become resonances and decay within a finite time. Thus, for the existence of a stationary qubit, the tunnel connection between antidots should  be not too small.

A qubit is equivalent to spin 1/2 in a ``magnetic field''
\be
\mathbf B_{\rm qubit} = (\Delta_x,\Delta_y, W).
\label{B-qubit}
\ee
Most importantly, the ``magnetic field'' $\mathbf B_{\rm qubit}$ depends on the gate voltage and can be controlled in a purely electrical way. Note also that such qubits exist in all bands so that at high temperatures the temperature window contains an ensemble of  $N  \sim T L/v_{\rm F}$ qubits. These qubits can be coherently controlled using gates and  magnetic field.

{\bf Conclusions.}
In the present work, we have developed a theory of a tunable crystal formed by a periodic set of identical holes in a two-dimensional topological insulator. We have demonstrated that tunneling through such an array can be controlled both magnetically and purely electrically using various gate electrode configurations.

We have derived an effective Hamiltonian describing the topologically protected localized states that arise near the FJ in such a crystal. Specifically, we have shown that at the boundary between regions with different  magnetic fluxes, two localized states arise that form a qubit placed in an effective ``magnetic field'' $\mathbf B_{\rm qubit}$ (see Eq.~\eqref{B-qubit}].  Remarkably,  the effective field depends on the gate voltage so that the state of the qubit can be controlled purely electrically.

In a crystal with several boundaries between different regions, spatially separated qubits arise. By changing (adiabatically or abruptly) external parameters, for example, the gate voltage, on which $\mathbf B_{\rm qubit}$ depends, one can control a single qubit and create entangled states involving qubits located at different boundaries. Moreover, even at one boundary, an ensemble of $N \sim T L/v_{\rm F}$ topologically protected qubits can arise at relatively high temperatures, $T\gg v_{\rm F}/L,$ when the temperature window covers many bands of the crystal and, therefore, includes many massive Dirac cones. An analysis of quantum computing by an ensemble of energetically separated qubits in the AB interferometer was presented in our recent paper~\cite{Niyazov2020}. The generalization of the results of ~\cite{Niyazov2020} to a helical crystal is an interesting problem for the further development of HES-based interferometry. 

Before closing the paper,  we  outline several problems that could be considered in further research. The most interesting is the study of the role of the electron-electron interaction. This interaction can lead to several non-trivial effects. First, if this interaction is taken into account, edge reconstruction~\cite{Wang2017a} is possible for some models of the confining potential at the sample edge. In this case, more than two edge conducting states can occur under certain assumptions about the edge potential. In addition, the interaction leads to a renormalization of the scattering matrix, which describes the jumps between neighboring rings. Such a renormalization can dramatically change the $\alpha$ and $\beta$ parameters describing jumps between neighboring rings~\cite{Aristov2016}. Finally, the interaction leads to  dephasing of the helical states. It would be interesting to clarify how such dephasing would affect the qubit's relaxation time (the so-called $T_1$ time) and dephasing time ($T_2$ time). It is expected that at low temperatures, $T \ll \Delta_{\rm DP}$, the relaxation rate, $T_1^{-1}$, will be suppressed, by analogy with the superconducting case, when the temperature becomes below the superconducting gap, \cite{Maleyev1994} as opposed to dephasing, which is determined by virtual processes.
It is also interesting to analyze the effects associated with the crystal's imperfection, which leads
to random components in the transfer matrix. On a qualitative level, such effects should lead to the appearance of a finite localization length. The corresponding theoretical analysis is apparently similar to the analysis of localization in a one-dimensional spinless chain of scattering centers described by random transfer matrices   (see \cite{Perel1984,Dmitriev1989}).
An analysis of these important issues is beyond the scope of this paper and may be considered in future studies.

Thus, we have demonstrated the possibility of creating controllable qubits in a helical crystal. The most interesting applications of the obtained results in the field of quantum computing appear due to possibilities of purely electrical high-temperature control of qubits.

{\bf Funding.}
The work was funded by 
the Russian Science Foundation, Grant No. 20-12-00147-$\Pi$.  The work of R.N. was partially supported by the Theoretical Physics and Mathematics Advancement Foundation ``BASIS''.

\end{document}